\pdfoutput=1

\documentclass[aps,prd,twocolumn,showpacs,amsmath,amssymb,amsfonts,preprintnumbers]{revtex4}
\usepackage{CJK}
\usepackage{graphicx}
\usepackage{amsmath}

\newcounter{mysectnumber}
\newcommand{\mysection}[1]{\textbf{\arabic{mysectnumber}.\stepcounter{mysectnumber}} \textsl{#1}. --}

\begin{document}

\title{Self-induced suppression of collective neutrino oscillations in a supernova}

\author{Huaiyu Duan}

\author{Alexander Friedland} \affiliation{Theoretical Division, MS B285, Los Alamos
  National Laboratory, Los Alamos, NM 87545, USA}

\date{Receive in PRL June 14, 2010; published March 2, 2011}

\begin{abstract}
We investigate collective flavor oscillations of supernova neutrinos at late stages of the explosion. We first show that the frequently used single-angle (averaged coupling) approximation predicts oscillations close to, or perhaps even inside, the neutrinosphere, potentially invalidating the basic neutrino transport paradigm. Fortunately, we also find that the single-angle approximation breaks down in this regime; in the full multiangle calculation, the oscillations start safely outside the transport region. The new suppression effect 
is traced to the interplay between the dispersion in the neutrino-neutrino interactions and the vacuum oscillation term. 

\end{abstract}

\preprint{LA-UR-10-03755}

\maketitle

\mysection{Introduction}
It was first suggested more than 20 years ago \cite{Fuller:1987aa,Notzold:1988kx} that interactions between neutrinos in a core-collapse supernova could impact their flavor evolution. Yet, modeling this effect has proven very challenging. 
Although recent years have brought considerable progress, we still do not have a complete picture of this phenomenon and surprises continue to emerge.

In its most general form, the problem involves modeling of a coupled
neutrino ensemble, with refraction and scattering, in an anisotropic
and fluctuating environment of an exploding supernova
(e.g., \cite{Strack:2005ux}). This is a formidable task, even for today's fastest supercomputers. Yet, it is also unnecessary if flavor transformations always happen safely outside of the neutrinosphere. In that case, the full quantum-kinetic equations could be replaced by the flavor oscillation equations for streaming neutrinos \cite{Pantaleone:1992eq,Pantaleone:1992xh}, which only keep the coherent (forward scattering) part of the full Hamiltonian \cite{Sigl:1992fn,McKellar:1992ja,Friedland:2003dv,Bell:2003mg,Friedland:2003eh,Friedland:2006ke}.
Modern analyses begin with this assumption and  check its consistency \emph{a posteriori}. 

Assuming additionally that the effect can be modeled in spherical symmetry, the problem becomes manageable, even though still computationally intensive: with ${\cal O}(10^{3})$ energy bins and ${\cal O}(10^{3})$ angular bins one needs to solve millions of coupled differential equations. This task was first tackled only in 2005, on a supercomputer \cite{Duan:2006an,Duan:2006jv}, revealing rich, unexpected physics: neutrinos leaving the ``self-coupling'' region undergo large flavor transformations and settle into characteristic nonthermal ``split'' spectra. Crucially, the transformations in \cite{Duan:2006an,Duan:2006jv,Fogli:2007bk} indeed occurred safely outside the neutrinosphere (starting at 70-80 km), providing a highly non-trivial consistency check for the free-streaming framework.

Are further simplifications of this problem possible? In this letter,
we will focus on what is known as the \emph{single-angle}
approximation \cite{Qian:1994wh,Pantaleone:1994ns}. In this
approximation, one assumes that 
neutrino wavefunctions are independent of neutrino
trajectories and that it is sufficient to follow the neutrino flavor
evolution along a single, representing trajectory.
The full calculation, as described above, is referred to as \emph{multiangle}.

In \cite{Duan:2006an},  the single-angle and multiangle numerical results were observed to be quite similar. The same observation was also later made in \cite{Fogli:2007bk}, where it was suggested as a general result. 
This came as a very welcome development for the field, as single-angle calculations allow one to quickly explore various regimes of collective oscillations. 
In recent years, this framework has been used to uncover more rich physics, for example, the interplay of collective oscillations with the usual MSW effect \cite{Duan:2007sh,Dasgupta:2008cd}, the possibility of multiple spectral splits \cite{Dasgupta:2009mg}, the appearance of the ``mixed'' spectra and three-flavor instabilities \cite{Friedland:2010sc, Dasgupta:2010cd} and other effects (see, e.g., \cite{Duan:2010bg} for a review).
With so much staked on the validity of the single-angle approximation, it is of great interest to establish any limitations it might have, and physical reasons for them.

\mysection{Setup of the calculations}
For reasons that will become clear shortly, we choose the conditions that occur several seconds into the explosion. We adopt the late-time spectra of
Ref.~\cite{Keil:2002in}, as previously used
in \cite{Dasgupta:2009mg,Friedland:2010sc} \footnote{Just as in \cite{Friedland:2010sc}, we
select the point $p=10$, $q=3.5$ from Table 6
in \cite{Keil:2002in}. It describes spectra of the form $\propto
E^{2}[1+\exp(E/T-\eta)]^{-1}$, with $\langle E_{\nu_{e}} \rangle=9.4$
MeV, $\langle E_{\bar\nu_{e}} \rangle=13.0$ MeV, $\langle
E_{\nu_{\mu,\tau}} \rangle=15.8$ MeV, and $L_{\nu_{e}}=4.1\times
10^{51}$ erg/s, $L_{\bar\nu_{e}}=4.3\times 10^{51}$ erg/s,
$L_{\nu_{\mu,\tau}}=7.9\times 10^{51}$ erg/s.} and take the density
profile of the neutrino-driven wind to be $\rho=\rho_{0}(10\mbox{
km}/r)^{3}$, with $\rho_{0}=10^{8}$ g/cm$^{-3}$, and $Y_{e}= 0.4$,
inspired by \cite{Arcones:2006uq}. For definiteness, we start by
choosing the inverted hierarchy (IH) of the neutrino masses, with $\theta_{13}=0.01$ and the standard oscillation parameters, as in \cite{Friedland:2010sc}. The neutrinosphere is modeled as a sharp sphere with $R=10$ km, without limb darkening \cite{Duan:2006an}. These ingredients represent a rather standard reference setup in the field. Assuming the consistency of the framework, the neutrinos are held unoscillated ``by hand'' for radial distances less than $r_\text{start}=40$ km (see later). We solve the multiangle equations (here and in what follows) using the same numerical code
FLAT \cite{Duan:2008eb} that was used for 3-flavor multiangle
calculations in \cite{Cherry:2010yc}.

\begin{figure}[t]
  \includegraphics[angle=0,width=\columnwidth]{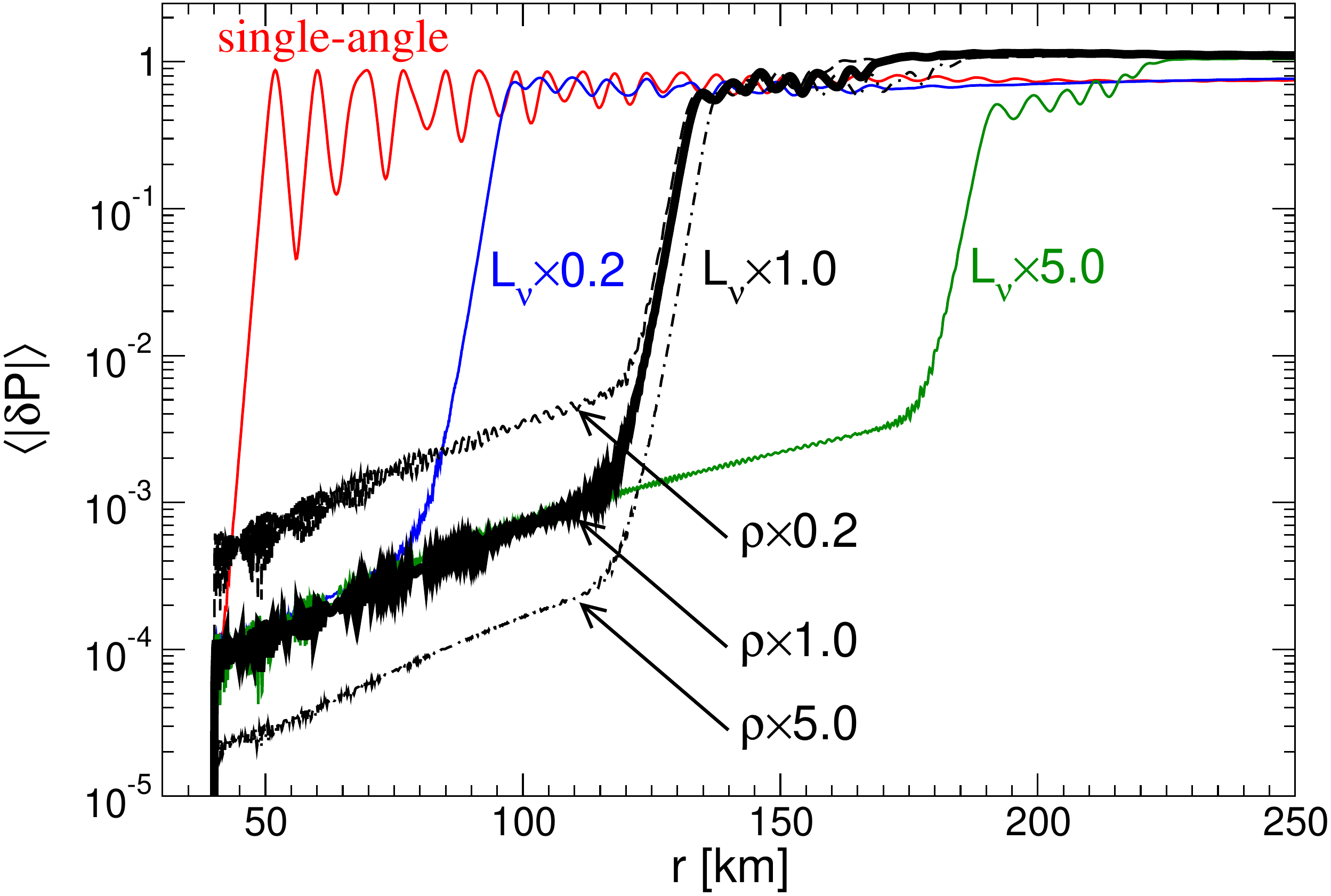}
  \caption{(Color online) The onset of the collective oscillations in the multiangle and single-angle cases. The onset is insensitive to the background density,  but depends strongly on the neutrino luminosities (as marked).}
  \label{fig:dP_3flavor}
\end{figure}

\mysection{Single- vs. multi-angle results}
The resulting evolution is shown in Fig.~\ref{fig:dP_3flavor}. Let us begin by considering the single-angle curve (marked ``single-angle''). The quantity plotted captures the deviation of the spectrum from the initial one: $|\langle \delta \boldsymbol{P}\rangle|\sim O(1)$ indicates large transformations, while $|\langle \delta \boldsymbol{P}\rangle|\ll1$ means the spectrum is nearly unchanged \footnote{The definition is
$|\langle \delta \boldsymbol{P} \rangle|=\int_{-\infty}^{\infty}dE_{\nu}\int d\Omega |\boldsymbol{P}(r)-\boldsymbol{P}(R)|/\int_{-\infty}^{\infty}dE_{\nu}\int d\Omega |\boldsymbol{P}(R)|$. Here, negative energies $E_{\nu}$ correspond to antineutrinos and $\boldsymbol{P}$ is the 8-dimensional $SU(3)$ ``polarization vector'', {\it i.e.}, the decomposition of the traceless part of the density matrix in the Gell-Mann basis \cite{Dasgupta:2007ws}. The norm is defined as a square root of the sum of the squares of the coefficients in the decomposition. }. A striking aspect of this result is that the neutrinos are unstable to oscillations the moment they are released. 
What this suggests is that the oscillations could be present close to the neutrinosphere. Indeed, repeating the calculations with smaller $r_\text{start}$, such as 30 km or even 20 km, we again find that the system is immediately unstable to oscillations. 
This finding is independent of the exact single-angle prescription: we found immediate oscillations whether we used the coupling for the radial trajectory, for the tangential, or for the one emitted at $45^{\circ}$ at the neutrinosphere. 
Thus, the short-distance suppression observed in the single-angle calculations in \cite{Duan:2006an,Duan:2006jv,Fogli:2007bk} is not a general result, but a consequence of the specific choice of conditions (emitted spectra) \footnote{The immediate instability was also present in the single-angle calculations in \cite{Friedland:2010sc} and, we believe, in \cite{Dasgupta:2009mg,Dasgupta:2010cd}, where similar initial spectra were used.}.

Taken at face value, this would mean neutrino transport has to be modeled together with collective oscillations. 
Fortunately, the multiangle calculations come to the rescue. They are also shown in Fig.~\ref{fig:dP_3flavor}, by the thick curve. In this case, the oscillations are \emph{suppressed} close to the protoneutron star. The flavor instability is seen to develop only at $r_{\rm inst}\sim120$ km. What we have, therefore, is a failure of the single-angle approximation and not of the neutrino transport paradigm.

\mysection{Deconstructing the effect}
Let us investigate the origin of this suppression. 
There are two conditions under which the single-angle approximation has been found to fail: (i) when the number fluxes of $\nu_{e}$ and $\bar\nu_{e}$ are highly symmetric \cite{Raffelt:2007yz,EstebanPretel:2007ec}, or (ii) in very dense matter \cite{EstebanPretel:2008ni,Duan:2008fd}. Neither of these conditions applies here. In particular, the high-density criterion is $n_{e^{-}}-n_{e^{+}}\gtrsim n_{\nu}$ \cite{EstebanPretel:2008ni}. 
This condition could be present during the initial stage of the explosion, when dense matter is piled up above the neutrinosphere. It should disappear, however, once the shock is pushed out. (This is the reason why we here focus on the late-time conditions.)

We can see explicitly that matter does not play a role if we increase and decrease the background density by a factor of five: as the dashed and dash-dotted curves in Fig.~\ref{fig:dP_3flavor} illustrate, the instability still sets in at $\sim 120$ km.

\begin{figure}[t]
  \includegraphics[angle=0,width=\columnwidth]{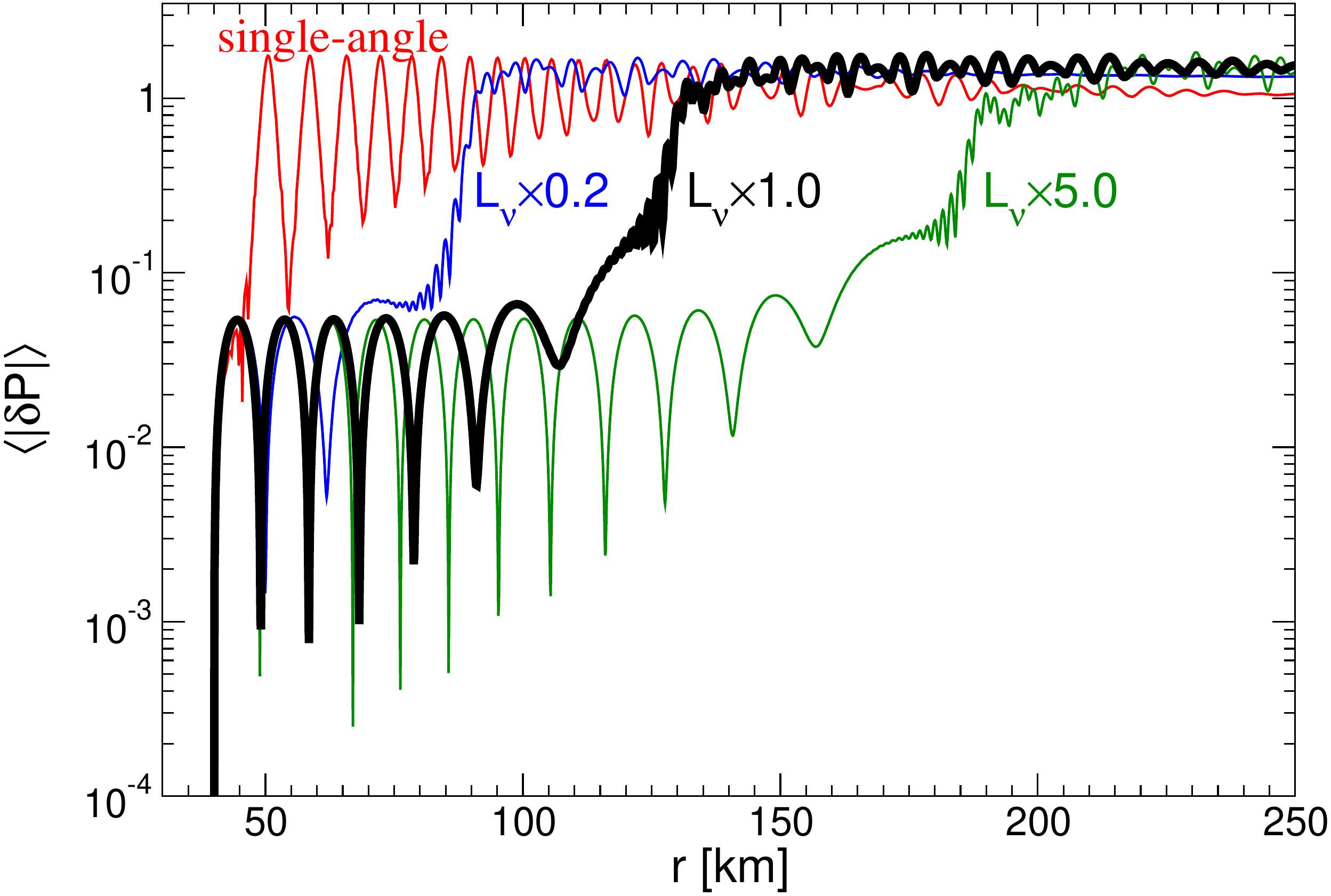}
  \caption{(Color online) Same as Fig.~\protect{\ref{fig:dP_3flavor}}, but with no matter and only two flavors. The onset of the oscillations and its dependence on the neutrino luminosities are virtually unchanged.}
  \label{fig:dP_2flavor}
\end{figure}

In fact, we can remove background matter altogether. 
In Fig.~\ref{fig:dP_2flavor} we show the results of the calculations for such a simplified setup. The thick curve shows that the large conversion still occurs around $r_{\rm inst}\sim120$ km. The calculation in this case is done with only two vacuum eigenstates separated by the atmospheric splitting, $\Delta m_\mathrm{atm}^{2}$. Thus, the three-flavor effects are not playing a crucial role in the location of the onset of the instability.

The only remaining ingredient in this simplified system is the $\nu-\nu$ interactions. These are different in the single- and multi-angle calculations: in the latter case, the strength of the potential felt by a neutrino depends on its direction. This \emph{dispersion} of the interaction strength must be the cause of the suppression.  

\mysection{Discussion}
First, let us review how dense matter suppresses collective modes.
In spherical symmetry, collective oscillations develop radially, so that on any radial shell
the oscillation phase is the same for all neutrinos. Yet, in the presence of matter and without $\nu-\nu$
coupling, neutrinos on different trajectories would accumulate
different phases between radial shells. Indeed, a neutrino
making an angle $\vartheta$ with the radial direction travels $\delta
r/\cos\vartheta$ between two spherical shells separated by $\delta
r$. 
(Unlike the common matter potential, 
this \emph{dispersion}, $\Delta H_{\rm mat}\sim H_{\rm
mat}(1/\cos\vartheta_{\rm max}-1)$, cannot be 
removed by using the corotating frame technique in \cite{Duan:2005cp}.) 
The $\nu-\nu$ interactions need to overcome $\Delta H_{\rm mat}$ to keep the neutrinos locked in a collective mode, resulting in the criterion $n_{e^{-}}-n_{e^{+}}\lesssim n_{\nu}$ \cite{EstebanPretel:2008ni,Duan:2008fd} for the oscillations.

It is easy to quantify how the potential felt by a neutrino depends on the angle $\vartheta$ its momentum makes with the radial direction. 
The interaction strength between a pair of neutrinos making an angle $\Theta$ is proportional to $(1-\cos\Theta)$. 
Integrating over the bundle of intersecting neutrino rays, one easily finds ({\it cf}. \cite{Pantaleone:1994ns,Duan:2006an})
\begin{equation}
	H_{\nu\nu} = \sqrt{2} G_{F} [n_{\nu_{e}}(r)-n_{\bar\nu_{e}}(r)]  
	[1-\cos\vartheta (1+x)/2],
	\label{eq:Hnunudispersion}
\end{equation}
where $x\equiv\sqrt{1-R^{2}/r^{2}}$ and $\cos\vartheta=\sqrt{1-\sin^{2}\vartheta_{R}R^{2}/r^{2}}$ in terms of the angle of emission $\vartheta_{R}$. In turn, the number densities are related to the fluxes $f_{i}=L_{i}/\langle E_{i}\rangle$ by
\begin{equation}
	n_{\nu_{i}}(r) = \frac{2 f_{i} (1-x)}{4\pi R^{2} c}.
\end{equation}

Strictly speaking, one should also include the geometric factor $1/\cos\vartheta$, just like dense matter. Its effect, however, appears only  at a higher order: for sufficiently large $r$, $H_{\nu\nu}\propto R^{2}/4r^{2}+\vartheta^{2}/2 + ...$, while the geometric correction appears only at $O(\vartheta^{4})$. This illustrates an important point: the dispersion in $\Delta H_{\rm mat}$
is small compared to $H_{\rm mat}$, while the dispersion in $ H_{\nu\nu}$ is \emph{of the same order as} $ H_{\nu\nu}$.  Therefore, the latter effect is more important, unless the matter is very dense.

To understand what sets the location of the instability $r_{\rm inst}$, let us first recall that in a single-angle calculation collective transformations are driven by the vacuum term $H_{\rm vac}=\Delta m^{2}/2 E_{\nu}$, or more accurately, by its \emph{dispersion} $\Delta H_{\rm vac}$ with energy (a common part can be removed, just like for the matter potential \cite{Duan:2005cp}). In the multiangle setup, the dispersion of $H_{\nu\nu}$ with angle $\vartheta$,
$\Delta H_{\nu\nu}\sim H_{\nu\nu}(\vartheta=\arccos x)-H_{\nu\nu}(\vartheta=0)$ adds in. Since $\Delta H_{\nu\nu}$ (just like $H_{\nu\nu}$) scales as $r^{-4}$, it dominates over  $\Delta H_{\rm vac}$ at small $r$. The regime when $\Delta H_{\nu\nu}$ ``overpowers'' $\Delta H_{\rm vac}$ is seen to give no oscillation. 
The flavor conversion therefore starts when the two become comparable,
\begin{equation}
 \Delta H_{\rm vac} \sim \Delta H_{\nu\nu}.
 \label{eq:condition}
\end{equation}
Plugging in the numbers, one indeed finds $r_{\rm inst}\sim120$ km. Furthermore, since $\Delta H_{\nu\nu}$ falls off as $r^{-4}$, while $H_{\rm vac}$ is independent of $r$, $r_{\rm inst}$ should depend on the luminosity as $L_{\nu}^{1/4}$. This is precisely what is seen in Figs.~\ref{fig:dP_3flavor} and \ref{fig:dP_2flavor}.

\begin{figure}[t]
  \includegraphics[width=\columnwidth]{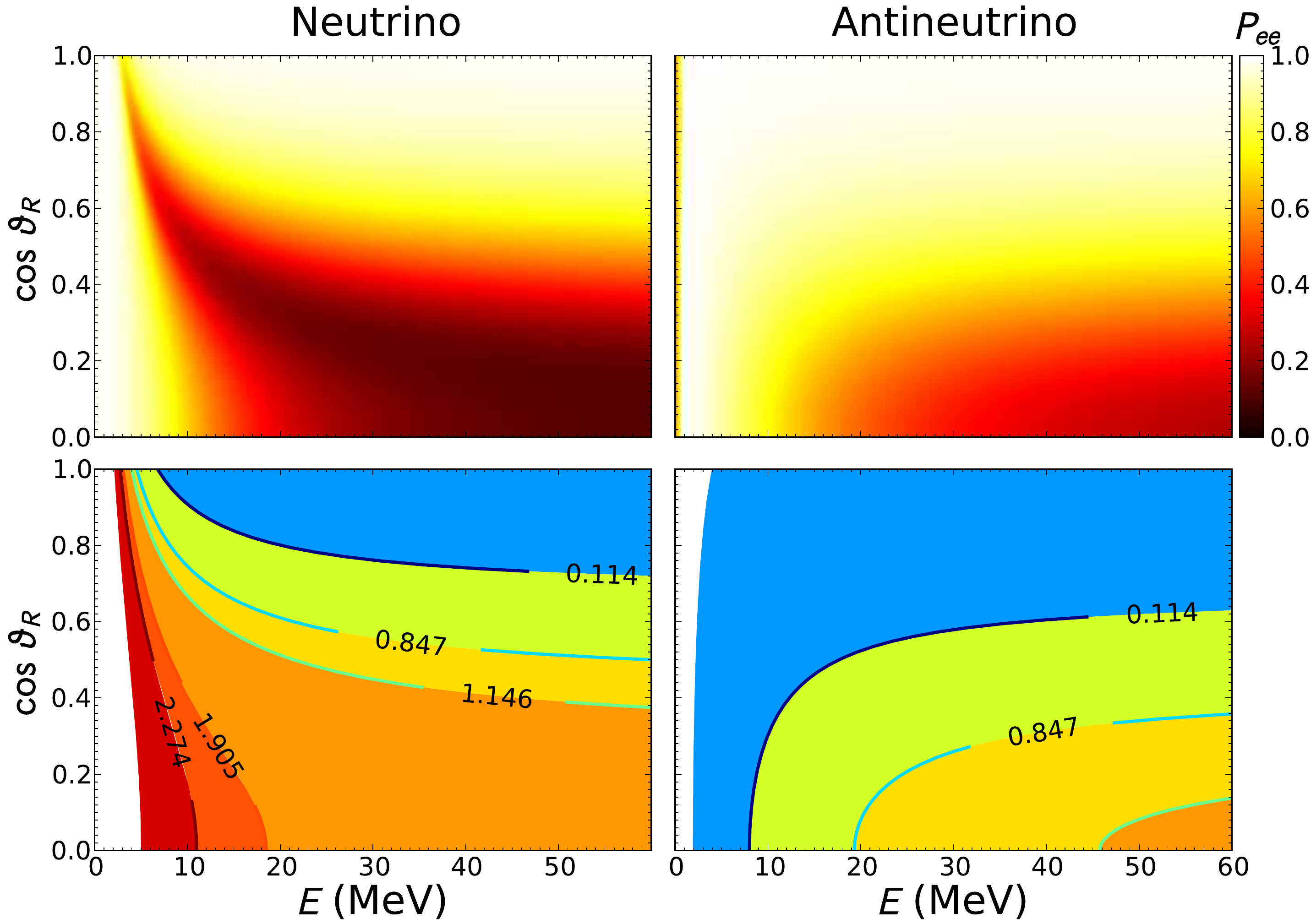}
  \caption{(Color online) \emph{(Top)}: The pattern of flavor
  transformations at the start of the oscillations, corresponding to
  the simulation shown by the thick curve in
  Fig.~\ref{fig:dP_3flavor}. \emph{(Bottom)}: Isocontours of
  $H_{\nu\nu}+H_{\rm vac} -(H_{\nu\nu}^{\rm max}+H_{\nu\nu}^{\rm
  min})/2 $, in units of $\Delta m_{\rm atm}^{2}/(20\mbox{
  MeV})$. Both are plotted for 
$r=133.6$ km.}
  \label{fig:thewedge}
\end{figure}

Further insight into this physics can be obtained from analyzing the oscillation mode in the $(E_{\nu},\vartheta_{R})$ plane. In Fig.~\ref{fig:thewedge}, in the top row we show the pattern of conversion at $r= 133.6$ km, where the instability just triggered conversion. As we have argued, the evolution at this point should be driven by a \emph{combination} of $\Delta H_{\rm vac}$ and $\Delta H_{\nu\nu}$. In the bottom rows of Fig.~\ref{fig:thewedge} we show the isocontours of $H_{\rm vac}+H_{\nu\nu}$. We can explicitly see that the two sources of the dispersion are comparable at this radius \footnote{If $\Delta H_{\rm vac}$ dominated, we would see vertical bands, while if $\Delta H_{\nu\nu}$ dominated the bands would be horizontal.} and furthermore, that \emph{the pattern of conversion matches the pattern of the isocontours remarkably well}.

We also investigated the normal mass hierarchy (NH) scenario. The oscillations in this case for our reference model also start at $r_{\rm inst}\sim120$ km. We confirmed that $r_{\rm inst}$ again scales with neutrino luminosity as $L^{1/4}$. The oscillation pattern in the $(E_{\nu},\vartheta_{R})$ plane is different from the IH case. Yet, once again, it follows the isocontours of $H_{\rm vac}+H_{\nu\nu}$ remarkably well. (In this case, the left and right panels of the bottom row in Fig.~\ref{fig:thewedge} switch places, because $\Delta m^{2}$ has the opposite sign.)

Animations depicting the flavor evolution as a function of $r$, for both IH and NH, are available online \footnote{http://prl.aps.org/supplemental/PRL/v106/i9/e091101; http://alexfriedland.com/papers/supernova/multiangle/; http://panda.unm.edu/\~{}duan/research/neutrino\_oscillations
/multiangle\_suppression/.}.

\begin{figure}[b]
  \includegraphics[width=\columnwidth]{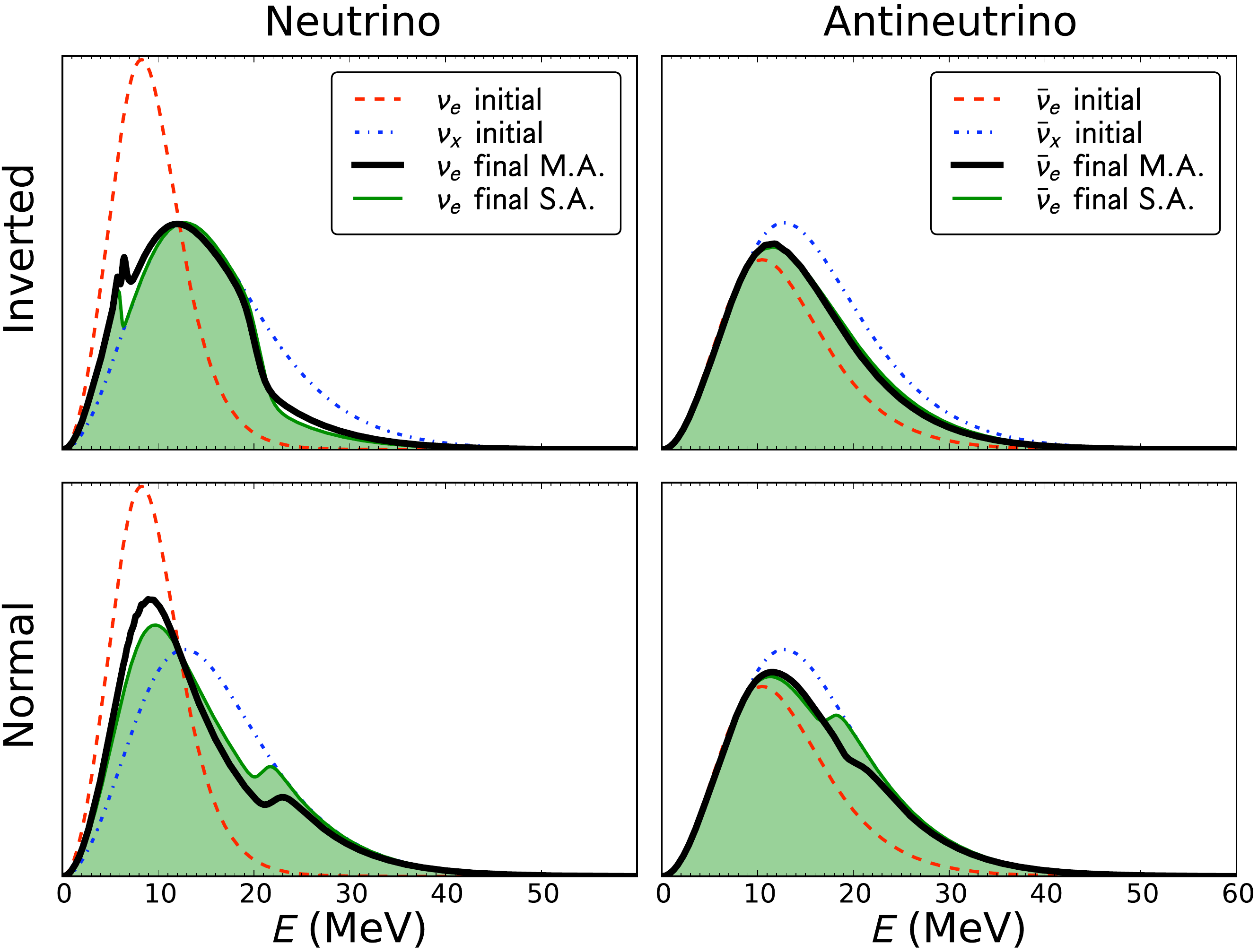}
   \caption{(Color online) Spectra of $\nu_{e}$ \emph{(left)} and
   $\bar\nu_{e}$ (\emph{right}) at infinity, for both
   IH \emph{(top)} and NH \emph{(bottom)}. Multiangle (M.A.) results are shown with thick solid
   curves,  single-angle (S.A.) with filled regions. Initial spectra are also shown, as marked. Turbulence \cite{Friedland:2006ta} and
   shock front \cite{Schirato:2002tg} effects are ignored.}
  \label{fig:finalspectra}
\end{figure}

\mysection{Final neutrino spectra}
In Fig.~\ref{fig:finalspectra}, we plot
the spectra ``at infinity'' for both neutrino mass hierarchies. Compared to the single-angle calculations, the multiangle ones give features that are somewhat smeared out, especially in the NH case, but are qualitatively similar \footnote{In comparing Fig.~\ref{fig:finalspectra} here to \cite{Friedland:2010sc}, one should keep in mind that the figures in \cite{Friedland:2010sc} show spectra for 1,000 km, before solar-scale MSW takes place.}. 
This behavior, however, is not general.  Our investigations show that for other spectra, the single- and multiangle calculations give qualitatively different answers not just at intermediate radii, but also at infinity. These results will be reported elsewhere.

\mysection{Generalizations} 
For applications,  one needs to know how the suppression effect manifests itself in a variety of conditions that may exist in a supernova. We have already seen that $r_{\rm inst}$ is insensitive to the matter profile, but varies with the overall neutrino luminosity. Another obvious factor is the radius of the neutrinosphere $R$, which decreases with time. Since $\Delta H_{\nu\nu}$ scales as $\vartheta_{\rm max}^{2}\propto R^{2}$, while also being proportional to $r^{-4}$, $r_{\rm inst}$ should be proportional to $\sqrt{R}$,  
\begin{equation}
	\label{eq:propto}
	r_{\rm inst} \propto R^{1/2} L^{1/4}.
\end{equation}
We have verified that doubling $R$ shifted  $r_{\rm inst}$ from 120 km to 170 km.

Less trivial is the dependence of the effect on the neutrino spectra. We have observed that, as the spectra are varied, the coefficient implicit in Eq.~(\ref{eq:condition}) and the starting oscillation pattern in the $(E_{\nu},\vartheta_{R})$  plane change. 
As already mentioned, for sufficiently different spectra (small $\nu_{x}$ fluxes), oscillations at small $r$ could be suppressed by the \emph{known single-angle} mechanism \cite{Duan:2005cp}. This was the regime studied the original multiangle calculations \cite{Duan:2006an} and in many subsequent papers. The study of different initial spectra is forthcoming.

\mysection{Conclusions}
We have described a new effect that suppresses collective neutrino oscillations close to the protoneutron star. The effect was traced to the dispersion of the neutrino-neutrino interaction strength. A physical condition behind this suppression was discussed.

The suppression has implications for supernova modeling: with collective flavor oscillations suppressed  at the neutrinosphere, it is not inconsistent to treat the transport of neutrinos separately from collective oscillations.
The suppression also impacts the nucleosynthesis yields, particularly the $r$-process, where the single-angle approximation gives a grossly inaccurate answer \cite{Duan:2010af}. 

The physics of collective flavor transformations continues to surprise us with its richness. As more conditions are investigated, a more complete physical understanding of the phenomenon will emerge. The role of numerical calculations in this field, as a tool of scientific discovery, has been indispensable.

We thank S. Reddy for useful feedback.
We gladly acknowledge the use of supercomputing resources at LANL through the Institutional Computing Program.
This work was supported by the DOE Office of Science and the LANL LDRD program.

\bibliography{multianglesplits}

\begin{thebibliography}{35}
\expandafter\ifx\csname natexlab\endcsname\relax\def\natexlab#1{#1}\fi
\expandafter\ifx\csname bibnamefont\endcsname\relax
  \def\bibnamefont#1{#1}\fi
\expandafter\ifx\csname bibfnamefont\endcsname\relax
  \def\bibfnamefont#1{#1}\fi
\expandafter\ifx\csname citenamefont\endcsname\relax
  \def\citenamefont#1{#1}\fi
\expandafter\ifx\csname url\endcsname\relax
  \def\url#1{\texttt{#1}}\fi
\expandafter\ifx\csname urlprefix\endcsname\relax\def\urlprefix{URL }\fi
\providecommand{\bibinfo}[2]{#2}
\providecommand{\eprint}[2][]{\url{#2}}

\bibitem[{\citenamefont{Fuller et~al.}(1987)\citenamefont{Fuller, Mayle,
  Wilson, and Schramm}}]{Fuller:1987aa}
\bibinfo{author}{\bibfnamefont{G.~M.} \bibnamefont{Fuller}},
  \bibinfo{author}{\bibfnamefont{R.~W.} \bibnamefont{Mayle}},
  \bibinfo{author}{\bibfnamefont{J.~R.} \bibnamefont{Wilson}},
  \bibnamefont{and} \bibinfo{author}{\bibfnamefont{D.~N.}
  \bibnamefont{Schramm}}, \bibinfo{journal}{Astrophys. J.}
  \textbf{\bibinfo{volume}{322}}, \bibinfo{pages}{795} (\bibinfo{year}{1987}).

\bibitem[{\citenamefont{N\"{o}tzold and Raffelt}(1988)}]{Notzold:1988kx}
\bibinfo{author}{\bibfnamefont{D.}~\bibnamefont{N\"{o}tzold}} \bibnamefont{and}
  \bibinfo{author}{\bibfnamefont{G.}~\bibnamefont{Raffelt}},
  \bibinfo{journal}{Nucl. Phys.} \textbf{\bibinfo{volume}{B307}},
  \bibinfo{pages}{924} (\bibinfo{year}{1988}).

\bibitem[{\citenamefont{Strack and Burrows}(2005)}]{Strack:2005ux}
\bibinfo{author}{\bibfnamefont{P.}~\bibnamefont{Strack}} \bibnamefont{and}
  \bibinfo{author}{\bibfnamefont{A.}~\bibnamefont{Burrows}},
  \bibinfo{journal}{Phys. Rev.} \textbf{\bibinfo{volume}{D71}},
  \bibinfo{pages}{093004} (\bibinfo{year}{2005}), \eprint{hep-ph/0504035}.

\bibitem[{\citenamefont{Pantaleone}(1992{\natexlab{a}})}]{Pantaleone:1992eq}
\bibinfo{author}{\bibfnamefont{J.~T.} \bibnamefont{Pantaleone}},
  \bibinfo{journal}{Phys. Lett.} \textbf{\bibinfo{volume}{B287}},
  \bibinfo{pages}{128} (\bibinfo{year}{1992}{\natexlab{a}}).

\bibitem[{\citenamefont{Pantaleone}(1992{\natexlab{b}})}]{Pantaleone:1992xh}
\bibinfo{author}{\bibfnamefont{J.~T.} \bibnamefont{Pantaleone}},
  \bibinfo{journal}{Phys. Rev.} \textbf{\bibinfo{volume}{D46}},
  \bibinfo{pages}{510} (\bibinfo{year}{1992}{\natexlab{b}}).

\bibitem[{\citenamefont{Sigl and Raffelt}(1993)}]{Sigl:1992fn}
\bibinfo{author}{\bibfnamefont{G.}~\bibnamefont{Sigl}} \bibnamefont{and}
  \bibinfo{author}{\bibfnamefont{G.}~\bibnamefont{Raffelt}},
  \bibinfo{journal}{Nucl. Phys.} \textbf{\bibinfo{volume}{B406}},
  \bibinfo{pages}{423} (\bibinfo{year}{1993}).

\bibitem[{\citenamefont{McKellar and Thomson}(1994)}]{McKellar:1992ja}
\bibinfo{author}{\bibfnamefont{B.~H.~J.} \bibnamefont{McKellar}}
  \bibnamefont{and} \bibinfo{author}{\bibfnamefont{M.~J.}
  \bibnamefont{Thomson}}, \bibinfo{journal}{Phys. Rev.}
  \textbf{\bibinfo{volume}{D49}}, \bibinfo{pages}{2710} (\bibinfo{year}{1994}).

\bibitem[{\citenamefont{Friedland and
  Lunardini}(2003{\natexlab{a}})}]{Friedland:2003dv}
\bibinfo{author}{\bibfnamefont{A.}~\bibnamefont{Friedland}} \bibnamefont{and}
  \bibinfo{author}{\bibfnamefont{C.}~\bibnamefont{Lunardini}},
  \bibinfo{journal}{Phys. Rev.} \textbf{\bibinfo{volume}{D68}},
  \bibinfo{pages}{013007} (\bibinfo{year}{2003}{\natexlab{a}}),
  \eprint{hep-ph/0304055}.

\bibitem[{\citenamefont{Bell et~al.}(2003)\citenamefont{Bell, Rawlinson, and
  Sawyer}}]{Bell:2003mg}
\bibinfo{author}{\bibfnamefont{N.~F.} \bibnamefont{Bell}},
  \bibinfo{author}{\bibfnamefont{A.~A.} \bibnamefont{Rawlinson}},
  \bibnamefont{and} \bibinfo{author}{\bibfnamefont{R.~F.}
  \bibnamefont{Sawyer}}, \bibinfo{journal}{Phys. Lett.}
  \textbf{\bibinfo{volume}{B573}}, \bibinfo{pages}{86} (\bibinfo{year}{2003}),
  \eprint{hep-ph/0304082}.

\bibitem[{\citenamefont{Friedland and
  Lunardini}(2003{\natexlab{b}})}]{Friedland:2003eh}
\bibinfo{author}{\bibfnamefont{A.}~\bibnamefont{Friedland}} \bibnamefont{and}
  \bibinfo{author}{\bibfnamefont{C.}~\bibnamefont{Lunardini}},
  \bibinfo{journal}{JHEP} \textbf{\bibinfo{volume}{10}}, \bibinfo{pages}{043}
  (\bibinfo{year}{2003}{\natexlab{b}}), \eprint{hep-ph/0307140}.

\bibitem[{\citenamefont{Friedland et~al.}(2006)\citenamefont{Friedland,
  McKellar, and Okuniewicz}}]{Friedland:2006ke}
\bibinfo{author}{\bibfnamefont{A.}~\bibnamefont{Friedland}},
  \bibinfo{author}{\bibfnamefont{B.~H.~J.} \bibnamefont{McKellar}},
  \bibnamefont{and}
  \bibinfo{author}{\bibfnamefont{I.}~\bibnamefont{Okuniewicz}},
  \bibinfo{journal}{Phys. Rev.} \textbf{\bibinfo{volume}{D73}},
  \bibinfo{pages}{093002} (\bibinfo{year}{2006}), \eprint{hep-ph/0602016}.

\bibitem[{\citenamefont{Duan et~al.}(2006{\natexlab{a}})\citenamefont{Duan,
  Fuller, Carlson, and Qian}}]{Duan:2006an}
\bibinfo{author}{\bibfnamefont{H.}~\bibnamefont{Duan}},
  \bibinfo{author}{\bibfnamefont{G.~M.} \bibnamefont{Fuller}},
  \bibinfo{author}{\bibfnamefont{J.}~\bibnamefont{Carlson}}, \bibnamefont{and}
  \bibinfo{author}{\bibfnamefont{Y.-Z.} \bibnamefont{Qian}},
  \bibinfo{journal}{Phys. Rev.} \textbf{\bibinfo{volume}{D74}},
  \bibinfo{pages}{105014} (\bibinfo{year}{2006}{\natexlab{a}}),
  \eprint{astro-ph/0606616}.

\bibitem[{\citenamefont{Duan et~al.}(2006{\natexlab{b}})\citenamefont{Duan,
  Fuller, Carlson, and Qian}}]{Duan:2006jv}
\bibinfo{author}{\bibfnamefont{H.}~\bibnamefont{Duan}},
  \bibinfo{author}{\bibfnamefont{G.~M.} \bibnamefont{Fuller}},
  \bibinfo{author}{\bibfnamefont{J.}~\bibnamefont{Carlson}}, \bibnamefont{and}
  \bibinfo{author}{\bibfnamefont{Y.-Z.} \bibnamefont{Qian}},
  \bibinfo{journal}{Phys. Rev. Lett.} \textbf{\bibinfo{volume}{97}},
  \bibinfo{pages}{241101} (\bibinfo{year}{2006}{\natexlab{b}}),
  \eprint{astro-ph/0608050}.

\bibitem[{\citenamefont{Fogli et~al.}(2007)\citenamefont{Fogli, Lisi, Marrone,
  and Mirizzi}}]{Fogli:2007bk}
\bibinfo{author}{\bibfnamefont{G.~L.} \bibnamefont{Fogli}},
  \bibinfo{author}{\bibfnamefont{E.}~\bibnamefont{Lisi}},
  \bibinfo{author}{\bibfnamefont{A.}~\bibnamefont{Marrone}}, \bibnamefont{and}
  \bibinfo{author}{\bibfnamefont{A.}~\bibnamefont{Mirizzi}},
  \bibinfo{journal}{JCAP} \textbf{\bibinfo{volume}{0712}}, \bibinfo{pages}{010}
  (\bibinfo{year}{2007}), \eprint{arXiv:0707.1998 [hep-ph]}.

\bibitem[{\citenamefont{Qian and Fuller}(1995)}]{Qian:1994wh}
\bibinfo{author}{\bibfnamefont{Y.~Z.} \bibnamefont{Qian}} \bibnamefont{and}
  \bibinfo{author}{\bibfnamefont{G.~M.} \bibnamefont{Fuller}},
  \bibinfo{journal}{Phys. Rev.} \textbf{\bibinfo{volume}{D51}},
  \bibinfo{pages}{1479} (\bibinfo{year}{1995}), \eprint{astro-ph/9406073}.

\bibitem[{\citenamefont{Pantaleone}(1995)}]{Pantaleone:1994ns}
\bibinfo{author}{\bibfnamefont{J.~T.} \bibnamefont{Pantaleone}},
  \bibinfo{journal}{Phys. Lett.} \textbf{\bibinfo{volume}{B342}},
  \bibinfo{pages}{250} (\bibinfo{year}{1995}), \eprint{astro-ph/9405008}.

\bibitem[{\citenamefont{Duan et~al.}(2008{\natexlab{a}})\citenamefont{Duan,
  Fuller, Carlson, and Qian}}]{Duan:2007sh}
\bibinfo{author}{\bibfnamefont{H.}~\bibnamefont{Duan}},
  \bibinfo{author}{\bibfnamefont{G.~M.} \bibnamefont{Fuller}},
  \bibinfo{author}{\bibfnamefont{J.}~\bibnamefont{Carlson}}, \bibnamefont{and}
  \bibinfo{author}{\bibfnamefont{Y.-Z.} \bibnamefont{Qian}},
  \bibinfo{journal}{Phys. Rev. Lett.} \textbf{\bibinfo{volume}{100}},
  \bibinfo{pages}{021101} (\bibinfo{year}{2008}{\natexlab{a}}),
  \eprint{0710.1271}.

\bibitem[{\citenamefont{Dasgupta et~al.}(2008)\citenamefont{Dasgupta, Dighe,
  Mirizzi, and Raffelt}}]{Dasgupta:2008cd}
\bibinfo{author}{\bibfnamefont{B.}~\bibnamefont{Dasgupta}},
  \bibinfo{author}{\bibfnamefont{A.}~\bibnamefont{Dighe}},
  \bibinfo{author}{\bibfnamefont{A.}~\bibnamefont{Mirizzi}}, \bibnamefont{and}
  \bibinfo{author}{\bibfnamefont{G.~G.} \bibnamefont{Raffelt}},
  \bibinfo{journal}{Phys. Rev.} \textbf{\bibinfo{volume}{D77}},
  \bibinfo{pages}{113007} (\bibinfo{year}{2008}), \eprint{0801.1660}.

\bibitem[{\citenamefont{Dasgupta et~al.}(2009)\citenamefont{Dasgupta, Dighe,
  Raffelt, and Smirnov}}]{Dasgupta:2009mg}
\bibinfo{author}{\bibfnamefont{B.}~\bibnamefont{Dasgupta}},
  \bibinfo{author}{\bibfnamefont{A.}~\bibnamefont{Dighe}},
  \bibinfo{author}{\bibfnamefont{G.~G.} \bibnamefont{Raffelt}},
  \bibnamefont{and} \bibinfo{author}{\bibfnamefont{A.~Y.}
  \bibnamefont{Smirnov}}, \bibinfo{journal}{Phys. Rev. Lett.}
  \textbf{\bibinfo{volume}{103}}, \bibinfo{pages}{051105}
  (\bibinfo{year}{2009}), \eprint{0904.3542}.

\bibitem[{\citenamefont{Friedland}(2010)}]{Friedland:2010sc}
\bibinfo{author}{\bibfnamefont{A.}~\bibnamefont{Friedland}},
  \bibinfo{journal}{Phys. Rev. Lett.} \textbf{\bibinfo{volume}{104}},
  \bibinfo{pages}{191102} (\bibinfo{year}{2010}), \eprint{1001.0996}.

\bibitem[{\citenamefont{Dasgupta et~al.}(2010)\citenamefont{Dasgupta, Mirizzi,
  Tamborra, and Tomas}}]{Dasgupta:2010cd}
\bibinfo{author}{\bibfnamefont{B.}~\bibnamefont{Dasgupta}},
  \bibinfo{author}{\bibfnamefont{A.}~\bibnamefont{Mirizzi}},
  \bibinfo{author}{\bibfnamefont{I.}~\bibnamefont{Tamborra}}, \bibnamefont{and}
  \bibinfo{author}{\bibfnamefont{R.}~\bibnamefont{Tomas}}
  (\bibinfo{year}{2010}), \eprint{1002.2943}.

\bibitem[{\citenamefont{Duan et~al.}(2010)\citenamefont{Duan, Fuller, and
  Qian}}]{Duan:2010bg}
\bibinfo{author}{\bibfnamefont{H.}~\bibnamefont{Duan}},
  \bibinfo{author}{\bibfnamefont{G.~M.} \bibnamefont{Fuller}},
  \bibnamefont{and} \bibinfo{author}{\bibfnamefont{Y.-Z.} \bibnamefont{Qian}}
  (\bibinfo{year}{2010}), \eprint{1001.2799}.

\bibitem[{\citenamefont{Keil et~al.}(2003)\citenamefont{Keil, Raffelt, and
  Janka}}]{Keil:2002in}
\bibinfo{author}{\bibfnamefont{M.~T.} \bibnamefont{Keil}},
  \bibinfo{author}{\bibfnamefont{G.~G.} \bibnamefont{Raffelt}},
  \bibnamefont{and} \bibinfo{author}{\bibfnamefont{H.-T.} \bibnamefont{Janka}},
  \bibinfo{journal}{Astrophys. J.} \textbf{\bibinfo{volume}{590}},
  \bibinfo{pages}{971} (\bibinfo{year}{2003}), \eprint{astro-ph/0208035}.

\bibitem[{\citenamefont{Arcones et~al.}(2007)\citenamefont{Arcones, Janka, and
  Scheck}}]{Arcones:2006uq}
\bibinfo{author}{\bibfnamefont{A.}~\bibnamefont{Arcones}},
  \bibinfo{author}{\bibfnamefont{H.-T.} \bibnamefont{Janka}}, \bibnamefont{and}
  \bibinfo{author}{\bibfnamefont{L.}~\bibnamefont{Scheck}},
  \bibinfo{journal}{Astron. Astrophys.} \textbf{\bibinfo{volume}{467}},
  \bibinfo{pages}{1227} (\bibinfo{year}{2007}), \eprint{astro-ph/0612582}.

\bibitem[{\citenamefont{Duan et~al.}(2008{\natexlab{b}})\citenamefont{Duan,
  Fuller, and Carlson}}]{Duan:2008eb}
\bibinfo{author}{\bibfnamefont{H.}~\bibnamefont{Duan}},
  \bibinfo{author}{\bibfnamefont{G.~M.} \bibnamefont{Fuller}},
  \bibnamefont{and} \bibinfo{author}{\bibfnamefont{J.}~\bibnamefont{Carlson}},
  \bibinfo{journal}{Comput. Sci. Disc.} \textbf{\bibinfo{volume}{1}},
  \bibinfo{pages}{015007} (\bibinfo{year}{2008}{\natexlab{b}}),
  \eprint{0803.3650}.

\bibitem[{\citenamefont{Cherry et~al.}(2010)\citenamefont{Cherry, Fuller,
  Carlson, Duan, and Qian}}]{Cherry:2010yc}
\bibinfo{author}{\bibfnamefont{J.~F.} \bibnamefont{Cherry}},
  \bibinfo{author}{\bibfnamefont{G.~M.} \bibnamefont{Fuller}},
  \bibinfo{author}{\bibfnamefont{J.}~\bibnamefont{Carlson}},
  \bibinfo{author}{\bibfnamefont{H.}~\bibnamefont{Duan}}, \bibnamefont{and}
  \bibinfo{author}{\bibfnamefont{Y.-Z.} \bibnamefont{Qian}}
  (\bibinfo{year}{2010}), \eprint{1006.2175}.

\bibitem[{\citenamefont{Raffelt and Sigl}(2007)}]{Raffelt:2007yz}
\bibinfo{author}{\bibfnamefont{G.~G.} \bibnamefont{Raffelt}} \bibnamefont{and}
  \bibinfo{author}{\bibfnamefont{G.~G.~R.} \bibnamefont{Sigl}},
  \bibinfo{journal}{Phys. Rev.} \textbf{\bibinfo{volume}{D75}},
  \bibinfo{pages}{083002} (\bibinfo{year}{2007}), \eprint{hep-ph/0701182}.

\bibitem[{\citenamefont{Esteban-Pretel
  et~al.}(2007)\citenamefont{Esteban-Pretel, Pastor, Tomas, Raffelt, and
  Sigl}}]{EstebanPretel:2007ec}
\bibinfo{author}{\bibfnamefont{A.}~\bibnamefont{Esteban-Pretel}},
  \bibinfo{author}{\bibfnamefont{S.}~\bibnamefont{Pastor}},
  \bibinfo{author}{\bibfnamefont{R.}~\bibnamefont{Tomas}},
  \bibinfo{author}{\bibfnamefont{G.~G.} \bibnamefont{Raffelt}},
  \bibnamefont{and} \bibinfo{author}{\bibfnamefont{G.}~\bibnamefont{Sigl}},
  \bibinfo{journal}{Phys. Rev.} \textbf{\bibinfo{volume}{D76}},
  \bibinfo{pages}{125018} (\bibinfo{year}{2007}), \eprint{arXiv:0706.2498
  [astro-ph]}.

\bibitem[{\citenamefont{Esteban-Pretel et~al.}(2008)}]{EstebanPretel:2008ni}
\bibinfo{author}{\bibfnamefont{A.}~\bibnamefont{Esteban-Pretel}}
  \bibnamefont{et~al.}, \bibinfo{journal}{Phys. Rev.}
  \textbf{\bibinfo{volume}{D78}}, \bibinfo{pages}{085012}
  (\bibinfo{year}{2008}), \eprint{0807.0659}.

\bibitem[{\citenamefont{Duan et~al.}(2009)\citenamefont{Duan, Fuller, and
  Qian}}]{Duan:2008fd}
\bibinfo{author}{\bibfnamefont{H.}~\bibnamefont{Duan}},
  \bibinfo{author}{\bibfnamefont{G.~M.} \bibnamefont{Fuller}},
  \bibnamefont{and} \bibinfo{author}{\bibfnamefont{Y.-Z.} \bibnamefont{Qian}},
  \bibinfo{journal}{J. Phys. G} \textbf{\bibinfo{volume}{36}},
  \bibinfo{pages}{105003} (\bibinfo{year}{2009}), \eprint{0808.2046}.

\bibitem[{\citenamefont{Duan et~al.}(2006{\natexlab{c}})\citenamefont{Duan,
  Fuller, and Qian}}]{Duan:2005cp}
\bibinfo{author}{\bibfnamefont{H.}~\bibnamefont{Duan}},
  \bibinfo{author}{\bibfnamefont{G.~M.} \bibnamefont{Fuller}},
  \bibnamefont{and} \bibinfo{author}{\bibfnamefont{Y.-Z.} \bibnamefont{Qian}},
  \bibinfo{journal}{Phys. Rev.} \textbf{\bibinfo{volume}{D74}},
  \bibinfo{pages}{123004} (\bibinfo{year}{2006}{\natexlab{c}}),
  \eprint{astro-ph/0511275}.

\bibitem[{\citenamefont{Friedland and Gruzinov}(2006)}]{Friedland:2006ta}
\bibinfo{author}{\bibfnamefont{A.}~\bibnamefont{Friedland}} \bibnamefont{and}
  \bibinfo{author}{\bibfnamefont{A.}~\bibnamefont{Gruzinov}}
  (\bibinfo{year}{2006}), \eprint{astro-ph/0607244}.

\bibitem[{\citenamefont{Schirato and Fuller}(2002)}]{Schirato:2002tg}
\bibinfo{author}{\bibfnamefont{R.~C.} \bibnamefont{Schirato}} \bibnamefont{and}
  \bibinfo{author}{\bibfnamefont{G.~M.} \bibnamefont{Fuller}}
  (\bibinfo{year}{2002}), \eprint{astro-ph/0205390}.

\bibitem[{\citenamefont{Duan et~al.}(2011)\citenamefont{Duan, Friedland,
  McLaughlin, and Surman}}]{Duan:2010af}
\bibinfo{author}{\bibfnamefont{H.}~\bibnamefont{Duan}},
  \bibinfo{author}{\bibfnamefont{A.}~\bibnamefont{Friedland}},
  \bibinfo{author}{\bibfnamefont{G.~C.} \bibnamefont{McLaughlin}},
  \bibnamefont{and} \bibinfo{author}{\bibfnamefont{R.}~\bibnamefont{Surman}},
  \bibinfo{journal}{J. Phys. G: Nucl. Part. Phys.}
  \textbf{\bibinfo{volume}{38}}, \bibinfo{pages}{035201}
  (\bibinfo{year}{2011}), \eprint{arXiv:1012.0532}.

\bibitem[{\citenamefont{Dasgupta and Dighe}(2008)}]{Dasgupta:2007ws}
\bibinfo{author}{\bibfnamefont{B.}~\bibnamefont{Dasgupta}} \bibnamefont{and}
  \bibinfo{author}{\bibfnamefont{A.}~\bibnamefont{Dighe}},
  \bibinfo{journal}{Phys. Rev.} \textbf{\bibinfo{volume}{D77}},
  \bibinfo{pages}{113002} (\bibinfo{year}{2008}), \eprint{0712.3798}.

\end{thebibliography}

\end{document}